\pdfoutput=1
\documentclass[11pt,a4paper]{article}
\usepackage{times,latexsym}
\usepackage[T1]{fontenc}
\usepackage[utf8]{inputenc}
\DeclareUnicodeCharacter{2212}{\ensuremath{-}} 
\usepackage{amsmath}   
\usepackage{url}

\usepackage[acceptedWithA]{tacl2021v1}

\usepackage{booktabs}   
\usepackage{multirow}   
\usepackage{tikz}       
\usetikzlibrary{arrows.meta,positioning,fit,shapes.geometric,calc,decorations.pathreplacing}
\usepackage{amssymb}    
\usepackage{pifont}     
\usepackage{graphicx}   

\hypersetup{
  pdftitle={Does Document Structure Help Dense Retrieval? A Placebo-Controlled
            Ablation of Four Mechanisms Across Two Corpora},
  pdfauthor={Andrey Kuehlkamp, Priscila Correa Saboia Moreira, Samuel Rund},
  pdfsubject={Information Retrieval},
  pdfkeywords={dense retrieval, retrieval-augmented generation, document
               structure, chunking, placebo control, nDCG}
}

\begin{document}

\title{Does Document Structure Help Dense Retrieval? A Placebo-Controlled
Ablation of Four Mechanisms Across Two Corpora}

\author{
  \begin{tabular}{ccc}
    Andrey Kuehlkamp & Priscila Correa Saboia Moreira & Samuel Rund \\
    \texttt{akuehlka@nd.edu} & \texttt{pmoreira@nd.edu} & \texttt{srund@nd.edu}
  \end{tabular} \\[0.6em]
  Center for Research Computing, University of Notre Dame \\
  Notre Dame, Indiana, USA
}

\maketitle

\begin{abstract}
Retrieval-augmented generation systems increasingly rely on document-structure treatments: structure-aligned chunking, LLM-generated chunk contexts, heading-path metadata, and hierarchical two-stage retrieval. Separate studies support each on different corpora, embedders, and metrics, and none control for a shared confound: any text prepended to a chunk perturbs its embedding.

We present a mechanism-isolating ablation testing all four treatments under one protocol, matching chunk sizes across conditions and adding a semantically null placebo---heading paths that are structurally valid but shuffled across documents. We score retrieval with a coverage-aware nDCG and test four pre-registered contrasts via document-clustered bootstrap with Holm correction, on two distant corpora: 200 Wikipedia Featured Articles (951 queries) and 1,585 QASPER papers (4,303 questions).

Organization helps, and the cause is content, not tokens: structure-aligned chunks with real heading paths beat contextualized fixed windows (+0.022 / +0.012 cov-nDCG@10) and the placebo (+0.010 / +0.016). Naive two-stage hierarchical retrieval hurts (-0.033 / -0.015), traceable to first-stage section recall. Gold structure beats LLM-induced structure on Wikipedia but not on QASPER. Effects are small ($dz$ 0.06-0.11) but Holm-significant and consistent across corpora.
\end{abstract}

\section{Introduction}
Emerging systems that put large language models to work on the scientific literature begin by turning a flat PDF into a structured object — a hierarchy of sections and passages — before any downstream reading, extraction, or search. Alexandria \cite{moreira2026alexandria}, for instance, decomposes each paper into pages, sections, and machine-readable markdown as a first step toward a provenance-linked, queryable record of its findings. Inducing that structure is not free, and it encodes an implicit bet: that document structure pays off downstream. For retrieval — the component that determines which passages such a system ever surfaces — that bet is largely untested. Does an induced hierarchy actually help retrieval, and if so, which mechanism carries the benefit?

Dense retrieval over chunked documents is the core component of most retrieval-augmented generation (RAG) systems. A growing literature argues that document structure — the boundaries, headings, and hierarchy an author placed — should inform how systems chunk, embed, and search documents. Four families of structure-derived treatments each claim retrieval gains. (i) Structure-aligned chunking cuts at discourse boundaries instead of fixed token windows \cite{qu-etal-2025-semantic,shaukat2026systematic}. (ii) Chunk contextualization prepends an LLM-written blurb that situates each chunk in its document \cite{anthropic2024contextual}. Late chunking and contextual embeddings are architectural variants \cite{gunther2024late, morris2025contextual}. (iii) Structure-metadata injection prepends the chunk's native heading path. (iv) Hierarchical retrieval routes queries to sections or tree nodes before it ranks chunks \cite{sarthi2024raptor,liu-etal-2021-dense-hierarchical}.
Two problems keep these claims from being actionable. This paper attacks both at once.

\emph{The mechanism problem.} The four treatments appear in separate papers, on different corpora, with different embedders and metrics. As a result, the field cannot say which mechanism carries the benefit. Worse, published results disagree in sign. Semantic-chunking gains fail to replicate across tasks \cite{qu-etal-2025-semantic}. Contextualization helps in-corpus retrieval but hurts in-document retrieval \cite{zhou2026chunking}. Naive late chunking loses 21 nDCG points on an extractive benchmark \cite{conti2025context}. Flat retrieval matches or beats multi-stage retrieval in some studies \cite{laitenberger2025stronger,shah2026taxonomy}, while two-stage retrieval reports +12\% top-1 elsewhere \cite{liu-etal-2021-dense-hierarchical}. A practitioner cannot tell whether to cut differently, prepend text, or restructure retrieval. The practitioner also cannot tell whether the reported gains are artifacts of corpus and embedder choice \cite{bhat2025rethinking}.

\emph{The measurement problem.} The treatments themselves confound the measurement. Any prefix — a situating context, a heading path — changes the chunk's embedding. A gain over a bare baseline therefore never separates organizational content from generic token injection. No prior study we could verify runs a semantically null control. Structure-aligned chunks also differ from fixed windows in size distribution, which confounds boundary placement with chunk length \cite{bhat2025rethinking}. Evidence for one query frequently spans multiple chunks, and binary relevance metrics score that incoherently. Also, the headline method in the contextualization family rests on a vendor self-benchmark, not on peer-reviewed measurement \cite{anthropic2024contextual}.

We contribute a single controlled instrument that addresses both problems: an 11-condition ablation. The construction isolates every mechanism, and the design closes every confound we identified. We size-match the fixed-window baselines per document to the structure-aligned conditions. A placebo arm prepends heading paths that are structurally valid but shuffled across documents. The placebo has the same token count and format but null content, so we measure the token-injection confound instead of assuming it away. The primary metric, coverage-nDCG@10, extends nDCG with novelty-aware partial credit over multi-span evidence, in the lineage of $\alpha$-nDCG \cite{clarke2008novelty}. We pre-registered four contrasts on one fixed cell (primary metric, strong embedder, no reranker). We test them with a document-clustered bootstrap under Holm correction. The full grid runs on two deliberately distant corpora: 200 Wikipedia Featured Articles with 951 leakage-filtered synthetic queries, and 1,585 QASPER papers with 4,303 native reader questions. Every mechanism claim therefore carries an explicit cross-corpus replication status instead of resting on a single measurement. All LLM stages (structure induction, contextualization, query generation) ran on a local Gemma-4-26B model. The measured half of the pipeline is fully deterministic.

The findings, with their cross-corpus status:
\begin{enumerate}
    \item Organization helps dense retrieval — small but real, on both corpora. Structure-aligned chunks with heading-path prefixes beat contextualized fixed windows by +0.022 cov-nDCG@10 on Wikipedia-FA and +0.012 on QASPER. Both gains are Holm-significant. The dominant share of the gain comes from where the text is cut, not from what is prepended. On both corpora, the move from fixed windows to bare structure-aligned chunks (A0→A2) is worth two to four times the marginal heading-path step (A2→A3).
    \item The benefit is organizational content, not token injection. Real heading paths beat the placebo on both corpora (+0.010 Wikipedia-FA, +0.016 QASPER — the largest QASPER contrast). Shuffled paths buy nothing over no prefix at all, or actively hurt. To our knowledge, this is the first placebo-controlled measurement of structure-metadata injection.
    \item LLM chunk contextualization underperforms a simple cut at structure. Contextualized fixed windows (the Anthropic-style arm) rank below bare structure-aligned chunks on both corpora. Under realistic token budgets, where prefix tokens count against the context window, the prepended contexts are strictly harmful.
    \item Naive two-stage hierarchical retrieval hurts. Routing to sections before chunk ranking loses to flat search over the identical index on both corpora (-0.033, -0.015). Both losses are Holm-significant, and this is the only contrast the conservative secondary model also flags on Wikipedia-FA. The failure is diagnosable. First-stage section recall is 0.60-0.68 on the 200-document corpus and collapses to 0.17–0.20 on the 1,508-document corpus. On QASPER, a cross-encoder reranker recovers nearly all of the loss (-0.015 → -0.004).
    \item The value of gold structure is corpus-dependent. Gold heading trees beat LLM-induced trees on Wikipedia-FA (+0.013, Holm-significant) but not on QASPER (+0.001, null). On QASPER, induced structure is statistically as good as gold. We report both readings: the gold advantage does not generalize, and a 26B model's induced structure loses nothing measurable on scientific papers.
\end{enumerate}

The methodological contribution is inseparable from these results. Without the size control, finding (1) would be confounded. Without the placebo, finding (2) would be unmeasurable. Without cross-corpus replication, we would have reported the gold-structure advantage in finding (5) as general, and it is not. We offer the design as a template for mechanism claims in retrieval research. The study itself is evidence that the template changes conclusions.

\section{Related Work}

\subsection{Chunking strategy}

Fixed-size chunking fragments discourse \cite{merola2026reconstructing}. Structure-aligned alternatives can outperform it by large margins. A 36-method, 6-domain, 5-embedder comparison reports paragraph-group chunking at mean nDCG@5 $\approx$ 0.46, versus < 0.25 for fixed character windows \cite{shaukat2026systematic}. But the gains are unstable. \citeauthor{qu-etal-2025-semantic} \cite{qu-etal-2025-semantic} — the keystone non-replication result — find semantic chunking's advantages inconsistent across document retrieval, evidence retrieval, and generation, and not worth the compute. A follow-up analysis attributes the measured gains chiefly to synthetic corpora stitched from unrelated documents \cite{zhou2026chunking}. Chunk-size effects also interact with task and embedder, and they flip sign between models \cite{bhat2025rethinking}. This instability motivates our design rather than obstructing it. We hold corpus, embedder, and chunk size fixed, and we vary boundary placement alone (Section \ref{sec:controls}).

\subsection{Contextualization and context-aware embeddings}

Anthropic's Contextual Retrieval \cite{anthropic2024contextual} prepends a 50–100-token LLM-generated situating context to each chunk. It reports 35–49\% reductions in top-20 retrieval failure. Those figures come from an internal, non-peer-reviewed benchmark, which is itself part of the gap we address. Architectural variants reach the same goal without generation. Late chunking pools token embeddings after a full-document encoder pass \cite{gunther2024late}. Contextual document embeddings train the encoder to represent neighbors \cite{morris2025contextual}. The ConTEB benchmark and InSeNT training target the setting directly \cite{conti2025context}. Yet added context is not reliably helpful. Late chunking drops 21 nDCG@10 on CovidQA \cite{conti2025context}. Anthropic-style contexts produce only marginal gains in independent evaluation \cite{merola2026reconstructing}. Contextualization helps in-corpus retrieval while it hurts in-document retrieval \cite{zhou2026chunking}. Our A1 arm implements the Anthropic recipe verbatim: per-chunk LLM contexts on fixed windows. This fragile mechanism therefore competes directly against boundary placement and metadata injection under one metric.

\subsection{Structure-metadata injection}

Production RAG systems commonly prepend a chunk's native heading path — the document title and section trail, essentially a breadcrumb. But we could not verify any study that isolates this practice as its own treatment, separate from boundary placement and from LLM-generated context. It is the least-studied of the four mechanisms. It is also the site of our strongest novelty claims: the isolated measurement (A2 vs. A3) and the placebo control (A3 vs. P) that separates path content from token injection.

\subsection{Hierarchical retrieval}

RAPTOR builds a synthetic summary tree by recursive clustering and retrieves across its levels \cite{sarthi2024raptor}. Note that RAPTOR constructs the tree — it does not use the document's native hierarchy. Our G-conditions close that scope difference with real heading trees. Two-stage retrieval helps in some settings. Dense Hierarchical Retrieval reports up to +12\% absolute top-1 over flat DPR \cite{liu-etal-2021-dense-hierarchical}. But strong recent results go the other way. A flat retrieve-then-read baseline that preserves document order matches or beats multi-stage systems \cite{laitenberger2025stronger}. Topology-aware analyses find that RAPTOR-style semantics-first retrieval loses to flat baselines on structural tasks \cite{shah2026taxonomy}. The hierarchy that does help is parent–child reranking, not route-to-section \cite{singh2025hierarchical}. Our A4/G4 arms contribute a controlled data point to this debate. A diagnostic (first-stage section recall) localizes why routing loses (Section \ref{sec:naivehurts}).

\subsection{Evaluation methodology}

Coverage- and diversity-aware effectiveness measures descend from $\alpha$-nDCG \cite{clarke2008novelty}, which scores novel information rather than repeated relevance. Our coverage-nDCG applies the same principle at the character-span level, where one question's evidence spans several chunks (Section \ref{sec:metrics}). Placebo-style null controls are standard in medicine, and researchers increasingly urge them for ML ablations. None of the structure/context retrieval studies above includes one. Finally, non-replication is documented within this subfield \cite{qu-etal-2025-semantic, zhou2026chunking},. We respond in kind: replication across corpora is part of the result, not future work.

\section{Study Design}

\subsection{Conditions and pre-registered contrasts}

All conditions share one corpus, one embedder per run, and one retrieval budget. They differ only in (a) where chunk boundaries fall, (b) what prefix precedes the content at embedding time, and (c) whether retrieval is flat or two-stage. Chunks are contiguous, non-overlapping spans that exactly tile each document. The prefix, when present, renders as \verb|[Title › Section › … ] \n\n content|. Prefix tokens are tracked separately from content tokens throughout.

\begin{figure*}[t]
\centering
\resizebox{\textwidth}{!}{%
\begin{tikzpicture}[
  font=\footnotesize,
  >=Stealth,
  cutline/.style={densely dashed, gray},
  guide/.style={densely dotted, gray!70},
  chunk/.style={draw, rounded corners=1pt, minimum height=0.6cm, align=center, fill=gray!8},
  small box/.style={draw, rounded corners=1pt, align=center, fill=blue!6, inner sep=3pt},
  panel title/.style={font=\bfseries\footnotesize},
  arm label/.style={font=\scriptsize, anchor=west}
]

\begin{scope}[shift={(0,0)}]
  \node[panel title] (ta) at (0,2.15) {(a) Structure-aligned chunking};

  \draw[thick] (-3.2,0.85) rectangle (3.2,1.45);
  \foreach \x in {-1.07,1.07} {
    \draw[cutline] (\x,0.75) -- (\x,1.55);
  }
  \foreach \x in {-1.6,1.6} {
    \fill[gray!45] (\x,0.85) rectangle (\x+0.11,1.45);
    \node[font=\scriptsize] at (\x+0.055,1.68) {\S};
  }
  \node[arm label] at (3.35,1.15) {A0 (ignores \S)};

  \foreach \x in {-1.6,1.6} { \draw[guide] (\x,0.8) -- (\x,-0.15); }

  \draw[thick] (-3.2,-0.8) rectangle (3.2,-0.2);
  \foreach \x in {-1.6,1.6} {
    \fill[gray!45] (\x,-0.8) rectangle (\x+0.11,-0.2);
    \draw[black, thick] (\x,-0.9) -- (\x,-0.1);
  }
  \node[arm label] at (3.35,-0.5) {A2 (cuts at \S)};

  \node[font=\scriptsize,gray] at (0,-1.2) {one document (paragraph stream); shaded bands are section headings (\S)};
\end{scope}

\begin{scope}[shift={(11.0,0)}]
  \node[panel title] (tb) at (0,2.15) {(b) Chunk contextualization (A1)};

  \node[small box, fill=gray!8, minimum width=1.7cm] (doc) at (-2.4,1.5) {full document};
  \node[draw, diamond, aspect=2, fill=orange!25, font=\scriptsize] (llm) at (0,1.5) {LLM};
  \draw[->] (doc) -- (llm);

  \node[small box, minimum width=5.2cm, fill=orange!12] (ctx) at (0,0.35)
    {LLM-generated context ($\leq$80 tok)};
  \node[chunk, minimum width=5.2cm, below=0pt of ctx] (chb) {chunk content (fixed window)};
  \draw[->] (llm) -- (ctx);

  \node[draw, densely dashed, rounded corners=3pt, inner sep=4pt,
        fit=(ctx)(chb)] (unit) {};
  \node[font=\scriptsize, gray, below=0.08cm of unit] {prepended, then embedded as one unit};
\end{scope}

\begin{scope}[shift={(0,-4.4)}]
  \node[panel title] (tc) at (0,2.05) {(c) Structure-metadata injection};

  \node[arm label, text=green!50!black] at (-2.7,1.55) {\checkmark~A3: the chunk's real heading path};
  \node[small box, minimum width=5.4cm, fill=green!10] (bc1) at (0,1.15)
    {Doc Title \textrightarrow\ \S2 Methods \textrightarrow\ 2.1 Setup};
  \node[chunk, minimum width=5.4cm, below=0pt of bc1] (ch1) {chunk content};

  \node[arm label, text=red!60!black] at (-2.7,-0.35) {\ding{55}~P: shuffled path from another document};
  \node[small box, minimum width=5.4cm, fill=red!8] (bc2) at (0,-0.75)
    {Doc Title \textrightarrow\ \S5 Results \textrightarrow\ 3.2 Ablation};
  \node[chunk, minimum width=5.4cm, below=0pt of bc2] (ch2) {chunk content \emph{(identical)}};
\end{scope}

\begin{scope}[shift={(11.0,-4.4)}]
  \node[panel title] (td) at (0,2.05) {(d) Hierarchical retrieval (A4 vs.\ A3)};

  \node[small box, minimum width=1.0cm] (q1) at (-3.8,0.95) {query};
  \node[small box, fill=orange!10, minimum width=2.0cm] (s1) at (-1.5,0.95)
    {stage 1:\\section router\\(top 5 of $N$)};
  \node[small box, fill=orange!10, minimum width=2.4cm] (s2) at (1.35,0.95)
    {stage 2:\\rank chunks\\in routed sections};
  \node[small box, minimum width=1.0cm] (r1) at (3.7,0.95) {results};
  \draw[->] (q1) -- (s1);
  \draw[->] (s1) -- (s2);
  \draw[->] (s2) -- (r1);
  \node[font=\scriptsize, red!70!black, align=center, below=0.3cm of s1] (drop)
    {\ding{55}\ gold section\\dropped here};
  \draw[->, red!70!black, densely dashed] (s1.south) -- (drop.north);

  \node[small box, minimum width=1.0cm] (q2) at (-3.8,-1.15) {query};
  \node[small box, fill=blue!8, minimum width=4.6cm] (flat) at (0.3,-1.15)
    {flat search over \emph{all} chunks (A3)};
  \node[small box, minimum width=1.0cm] (r2) at (3.7,-1.15) {results};
  \draw[->] (q2) -- (flat);
  \draw[->] (flat) -- (r2);
  \node[font=\scriptsize, gray, align=center, below=0.12cm of flat]
    {(no routing step, so the gold section cannot be lost)};
\end{scope}

\end{tikzpicture}%
}
\caption{The four document-structure mechanisms under test. (a) Structure-aligned chunking cuts at section boundaries (A2) instead of ignoring them (A0). (b) Chunk contextualization prepends an LLM-generated blurb to a fixed-size window (A1). (c) Structure-metadata injection prepends the chunk's real heading path (A3); the placebo (P) prepends an equally-shaped but shuffled, semantically null path. (d) Hierarchical retrieval routes queries to sections before ranking chunks (A4); flat retrieval (A3) searches the full chunk index directly. A4 underperforms because stage 1 frequently drops the gold section before stage 2 ever sees it (Section \ref{sec:naivehurts}).}
\label{fig:mechanisms}
\end{figure*}
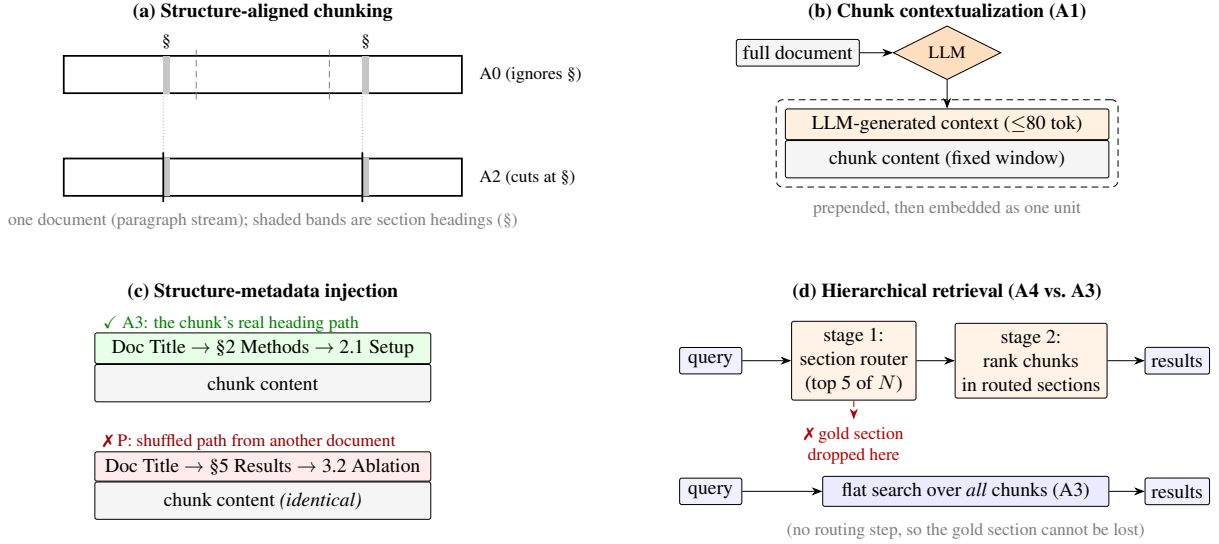

Figure \ref{fig:mechanisms} shows what each mechanism does to a document at embedding and retrieval time.

\subsection{Controls and design invariants}
\label{sec:controls}

We fixed four contrasts in the configuration before the runs. Each contrast isolates one mechanism question:
\begin{itemize}
    \item C1 (organization vs. contextualization), A3-A1: do structure-aligned boundaries plus native metadata beat the strongest generative treatment on fixed windows?
    \item C2 (content vs. tokens), A3-P: does the content of the heading path matter, or is the gain token injection?
    \item C3 (cost of induction), G3-A3: how much is lost when an LLM induces the structure instead of the author?
    \item C4 (routing vs. flat), A4-A3: does two-stage hierarchical retrieval beat flat search over the identical index?
\end{itemize}
Table \ref{tab:conditions-full} in the appendix lists the experimental conditions employed in this study.

\textbf{Size matching (guards C1).} The fixed-window baseline is not cut at an arbitrary size. Per document, its target window equals the mean token length of that document's induced chunks. Realized means differ by $\leq 4$ tokens per chunkset (Wikipedia-FA: fixed 255.4 vs. induced 259.6 content tokens; QASPER: 244.9 vs. 249.8). The comparison of A0/A1 vs. A2/A3 therefore isolates boundary placement, not chunk size. Window edges snap to sentence boundaries. Structural chunks split (at sentence boundaries) or merge when they exceed 350 or fall below 60 content tokens.

\textbf{Placebo construction (guards C2).} Placebo paths are real heading paths from other documents. We bucket unique (document, path) pairs by depth and permute them within-bucket toward a cross-document derangement (fixed seed). We count and report residual same-document assignments. Placebo prefixes therefore match real breadcrumbs in format, depth, and length distribution, while they carry no information about the chunk. This is a semantically null treatment in the strict sense.

\textbf{Representation constancy (guards C4).} The two-stage conditions build section centroids as the renormalized mean of the same chunk vectors that flat retrieval searches. C4 therefore compares retrieval algorithms, never representations. Stage 1 ranks section centroids and keeps the top 5 (a section = document + top-level heading). Stage 2 ranks chunks within the routed sections. If fewer than k candidates remain, the system backfills from the global ranking. A per-query section-recall diagnostic records whether stage 1 retained any section that overlaps the gold evidence. This separates routing failures from ranking failures.

\textbf{Leakage and provenance controls.} The pipeline selects the gold evidence spans. The query-writing LLM never self-reports them. Synthetic queries pass a two-round lexical-leakage filter (Section \ref{sec:metrics}). Relevance is computed purely from character-span overlap against content-only offsets, so prefixes can never create spurious relevance.

\subsection{Structure induction}

The induced-structure conditions (A2–A4, P) use a hierarchy from a local \emph{Gemma-4-26B-A4B} model (the Mixture-of-Experts variant) \cite{gemmateam2026gemma4}. The model sees the document's paragraphs with all headings stripped. It returns sections as (title, level, paragraph-range) tuples under hard constraints: sections tile the paragraphs exactly, and levels increase by at most one. We verify the constraints mechanically. A failure gets one self-correction round with the specific violations, then a deterministic repair that snaps near-valid output to a legal tiling. On Wikipedia-FA, 198/200 documents produced usable structure (116 required repair; mean 21.9 sections/doc). On QASPER, 1,579/1,585 did (590 repaired; mean 14.5 sections/doc vs. 14.8 gold). One fact is critical for C3: the inducer does not simply recover native sectioning. Only 78.9\% (Wikipedia-FA) and 75.8\% (QASPER) of induced chunk spans are byte-identical to a gold span. Gold and induced conditions therefore measure genuinely different segmentations.

\section{Corpora and Query Sets}

We chose the two corpora to be maximally distant while both satisfy the protocol's requirements: long documents with native hierarchical structure (so gold structure exists) and multi-span evidence (so coverage matters).

\subsection{Wikiperia Featured Articles (200 documents, synthetic queries)}

We sample documents from Wikipedia's Featured Articles under quality filters ($\geq 3,000$ words, heading depth $\geq 3$, $\geq 12$ sections) with topic-bucket balancing \footnote{\textit{We classify each article into one of six keyword-defined topic buckets (history/war, biography, science/nature, arts/media, geography/places, other) from its Wikipedia categories, and cap each bucket at 60 documents during seeded sampling (seed 20260710). The cap binds for two buckets; the realized corpus is history/war 60, other 60, biography 30, arts/media 22, science/nature 15, geography/places 13. The corpus remains topic-skewed, reflecting real Featured-Article imbalance; the cap limits rather than eliminates it.}}. A single canonical normalizer flattens the wikitext. Gold structure is the article's own heading tree, offset-verified.

Queries are LLM-generated but provenance- and leakage-controlled. For each document, the pipeline itself samples gold spans and requests four query types. A factoid query targets one span. An underspecified query targets one span but names the referent only vaguely — "the commander," never the proper name. A multihop query targets two spans in different top-level sections. A structural query targets a whole top-level section. Each generated query must pass a lexical-leakage filter: reject if ROUGE-L F1 vs. its gold span $> 0.35$ or if any verbatim 4-gram is shared. A second LLM call then paraphrases the query. The paraphrase — which is what enters the benchmark — must pass the same filter again. The funnel: 2,306 generated → 1,945 survivors → 951 balanced finals (250 factoid, 250 underspecified, 250 structural, 201 multihop) over 184 documents. Shipped queries are lexically far from their evidence (typical ROUGE-L 0.02–0.13). BM25 places the gold chunk first for only 36.3\% of queries, so the benchmark is not trivially lexical.

\subsection{QASPER (1,585 papers, native questions)}

QASPER \cite{dasigi2021dataset} provides NLP papers with questions from real readers and gold evidence paragraphs — human-authored relevance on a genre far from encyclopedic prose. We merge all splits (train/val/test), because the study is an Information Retrieval (IR) ablation, not supervised learning. Gold structure is the paper's own section tree. We harvest questions instead of generating them. We match each question's evidence paragraphs to the flattened document (exact → fuzzy → section fallback; 95.7\% exact). This yields 5,049 → 4,303 usable questions over 1,508 papers, after we drop unanswerable, no-evidence, and numeric-only items. The set is hard and non-lexical (BM25 gold-top-1 22.7\%). It is also genuinely multi-span: 43\% of questions cite multiple evidence paragraphs (mean 2.10 spans/question). That is precisely the regime coverage-nDCG exists for.

One structural property matters when we interpret C3. QASPER's native hierarchy is shallow: 61\% of papers have depth-1 section lists, versus depth $\geq 3$ for every Wikipedia-FA document by construction. Gold and induced trees therefore have less room to diverge (Section \ref{sec:corpus-struct}).

\section{Evaluation Protocol}

\subsection{Metrics}
\label{sec:metrics}

\textbf{Coverage-nDCG@10 (primary).} Walking down the ranking, a chunk's gain is the fraction of gold-evidence characters it covers that earlier chunks did not already cover. The DCG of these marginal gains is normalized by the DCG of the ideal ordering achievable from that condition's own chunk set. The novelty term prevents small-chunk conditions from banking the same span repeatedly. Per-condition normalization scores each condition against the best it could have done with its own segmentation, so a condition is not penalized merely for a different chunking. Both properties are required for fair comparison across chunkings. This is why binary nDCG is insufficient here.

\textbf{Secondary metrics} form a ladder from rank-insensitive to application-realistic: recall@50 over gold spans ($\geq 50\%$ covered by the union of top-50), Mean Reciprocal Rank (MRR), binary nDCG@10, and token-budget coverage at 512/1024/2048 tokens. In the token-budget metrics, \textbf{prefix tokens count against the budget}. Each treatment is charged its real context-window cost.

\textbf{Relevance rule.} A chunk is relevant to a span if and only if it covers $\geq 50\%$ of the span, or $\geq 80\%$ of the chunk lies inside the span. All overlap is computed on content-only character offsets.

\subsection{Statistical protocol}
\label{sec:statistical}

We test all four contrasts on one cell fixed in advance --- cov-nDCG@10, strong embedder, no reranker --- to prevent metric shopping. Each contrast is the unweighted mean of paired per-query deltas. Multiple queries share a document, so we estimate uncertainty with a cluster bootstrap that resamples documents with replacement (10,000 iterations) \cite{field2007bootstrapping}. P-values are two-sided and Holm-corrected across the four contrasts \cite{holm1979simple}. Effect sizes are Cohen's $dz$ \cite{lakens2013calculating}. Everything outside the primary cell is explicitly exploratory, and Section \ref{sec:results} labels it as such. This includes per-condition means across metrics and cells, the breadcrumb-depth dose-response, and per-query-type breakdowns.

We report a mixed-effects regression (score ~ condition, random intercept per document) as a secondary check. We flag its specification honestly. It omits a per-query random effect although treatment varies within query. Query-level variance therefore inflates its standard errors, and the model is conservative for small paired effects. We report it because pre-registration said we would. Discordances are discussed where they occur (Section \ref{sec:results-contrasts}).

\subsection{Implementation}

Embedders are \emph{bge-small-en-v1.5} ("weak") and \emph{bge-large-en-v1.5} ("strong") \cite{xiao2024cpack}. Every run uses exactly one embedder. The pair exists to test robustness across encoder capacity, not as a two-stage system. The reranked cells apply \emph{bge-reranker-v2-m3} \cite{chen2024m3} over the top 50 candidates of the strong embedder. All generative stages (structure induction, A1 contexts, query generation and paraphrase) ran on local \emph{Gemma-4-26B-A4B} via vLLM \cite{kwon2023efficient} on a single H200 GPU. Generation is confined to one-time corpus and query construction under fixed seeds with on-disk caching. The measured half of the pipeline --- embedding, retrieval, scoring, statistics --- is deterministic. Code, configurations, and per-query results will be released.

\section{Results}
\label{sec:results}

\subsection{The four pre-registered contrasts}
\label{sec:results-contrasts}

Table \ref{tab:contrasts} reports the primary cell. Three contrasts replicate across both corpora with the same sign and Holm significance. One is corpus-dependent.

\begin{table*}[t]
\centering
\small
\caption{Pre-registered contrasts, cov-nDCG@10, strong embedder, no reranker. Bootstrap CIs and Holm-corrected $p$-values. MixedLM \cite{seabold2010statsmodels, seabold2025statsmodels_software} is the conservative secondary model (See sec. \ref{sec:statistical}).}
\label{tab:contrasts}
\begin{tabular}{@{}llccccc@{}}
\toprule
\textbf{Contrast} & \textbf{Corpus} & \boldmath{$\Delta$} & \textbf{95\% CI} & \textbf{$p$(Holm)} & \boldmath{$dz$} & \textbf{$p$(MixedLM)} \\
\midrule
\multirow{2}{*}{C1: A3$-$A1} & Wikipedia-FA & \textbf{+0.0218} & [+0.0052, +0.0380] & 0.017 & +0.09 & 0.120 \\
 & QASPER & \textbf{+0.0115} & [+0.0055, +0.0175] & $<$0.001 & +0.06 & 0.008 \\
\addlinespace
\multirow{2}{*}{C2: A3$-$P} & Wikipedia-FA & \textbf{+0.0099} & [+0.0024, +0.0173] & 0.017 & +0.08 & 0.487 \\
 & QASPER & \textbf{+0.0161} & [+0.0114, +0.0207] & $<$0.001 & +0.11 & $<$0.001 \\
\addlinespace
\multirow{2}{*}{C3: G3$-$A3} & Wikipedia-FA & \textbf{+0.0129} & [+0.0038, +0.0228] & 0.010 & +0.10 & 0.373 \\
 & QASPER & +0.0011 & [$-$0.0020, +0.0043] & 0.485 (n.s.) & +0.01 & 0.801 \\
\addlinespace
\multirow{2}{*}{C4: A4$-$A3} & Wikipedia-FA & \textbf{$-$0.0326} & [$-$0.0442, $-$0.0217] & $<$0.001 & $-$0.17 & 0.027 \\
 & QASPER & \textbf{$-$0.0147} & [$-$0.0192, $-$0.0104] & $<$0.001 & $-$0.11 & $<$0.001 \\
\bottomrule
\end{tabular}
\end{table*}

On Wikipedia-FA the secondary MixedLM flags only C4. This is the expected signature of its specification, not evidence that the other effects are fragile. The model lacks a query random effect, so it absorbs large query-to-query variance into its residual. It can therefore detect only the largest effect ($|dz| = 0.17$, roughly twice the others). At QASPER's $n = 4,303$ queries, even this conservative model confirms all three non-null contrasts. The bootstrap is correctly specified for a paired, document-clustered design, and it is primary throughout.

Effect sizes are small: +0.01-0.02 cov-nDCG@10, $dz$ 0.06-0.11. We ask the reader to hold both halves of that sentence. The effects are small, but they are also real: given tight CIs, Holm correction, and same-sign replication across two genres and two query provenances. Steady gains of this size are within the range IR practice treats as meaningful. In a subfield documented for non-replication, the contrasts' cross-corpus consistency is itself a finding.

\subsection{Where you cut matters more than what you prepend}
\label{sec:wherecut}

Table \ref{tab:per-condition} shows per-condition means decompose the organization effect (strong embedder, no reranker):

\begin{table*}[t]
\centering
\small
\caption{Per-condition cov-nDCG@10 (strong embedder, no reranker).}
\label{tab:per-condition}
\begin{tabular}{@{}lccccccccc@{}}
\toprule
\textbf{Corpus} & \textbf{A0} & \textbf{A1} & \textbf{A2} & \textbf{A3} & \textbf{P} & \textbf{A4} & \textbf{G2} & \textbf{G3} & \textbf{G4} \\
\midrule
Wikipedia-FA & 0.362 & 0.391 & 0.403 & 0.413 & 0.403 & 0.381 & 0.417 & 0.426 & 0.400 \\
QASPER       & 0.098 & 0.114 & 0.117 & 0.125 & 0.109 & 0.111 & 0.120 & 0.126 & 0.111 \\
\bottomrule
\end{tabular}
\end{table*}

Two regularities hold on both corpora. First, the boundary step dominates the metadata step. A0→A2 — a re-cut of the same text at induced structural boundaries, with no prefix --- is worth +0.041 on Wikipedia-FA and +0.019 on QASPER. A2→A3, the addition of the real heading path, adds only +0.011 and +0.009. Second, A1 $<$ A2. LLM-written per-chunk contexts on fixed windows lose to bare structural cutting with no prefix at all. \emph{The generative treatment does not recover what bad boundaries destroy.} Both regularities also hold on the weak embedder (Wikipedia-FA A0 0.306 → A3 0.343; QASPER 0.087 → 0.107). The ordering is therefore not an artifact of encoder capacity.

Contextualization's cost becomes explicit under token budgets, where prefixes are charged to the context window. At a 512-token budget, A1 falls below plain A0 on both corpora (Wikipedia-FA 0.167 vs. 0.187; QASPER 0.046 vs. 0.052). A1 never clearly recovers by 2,048 tokens (0.396 vs. 0.397). An 80-token generated context that mostly restates document-level information displaces retrievable evidence.

\subsection{Content, not tokens: the placebo verdict}

The placebo arm exists to answer one question: is the A2→A3 gain merely "more tokens move the embedding somewhere useful"? On both corpora the answer is no. Real paths beat shuffled paths of identical format, depth, and length (+0.0099 Wikipedia-FA; +0.0161 QASPER, where C2 is the largest and most significant contrast). The placebo's position against the bare structural arm tells the same story. Shuffled paths buy nothing over no prefix at all (Wikipedia-FA: P 0.403 vs. A2 0.403), or they actively hurt (QASPER: P 0.109 vs. A2 0.117). The full QASPER ordering is instructive: A0 0.098 $<$ P 0.109 $<$ A1 0.114 $<$ A2 0.117 $<$ A3 0.125. Null tokens buy a little over raw windows. Contextualization buys more. Bare structural cutting buys more still. Real metadata buys the most.

Without arm P, none of this would be attributable. A gain from any added prefix could be an embedding-geometry artifact, and the A2→A3 step would remain ambiguous between content and token injection. With arm P, the attribution is measured. The prefix's value depends on its being true of the chunk. Semantically null tokens are at best neutral and at worst noise.

\subsection{Naive hierarchical retrieval hurts, and we can say why}
\label{sec:naivehurts}

C4 is negative on both corpora. It is the most robust single effect in the study (Wikipedia-FA $dz = -0.17$). The two-stage conditions retrieve over the identical vectors as A3, so the penalty is attributable entirely to routing. The section-recall diagnostic localizes it. The probability that stage 1 even retains a gold-bearing section is only 0.60–0.68 on the 200-document Wikipedia-FA corpus. It collapses to 0.17–0.20 on QASPER, where the router faces $\approx 22,000$ candidate sections. Whatever stage 1 drops, stage 2 cannot recover. Recall@50 shows the same shadow (QASPER: A4 0.293 vs. A3 0.302; Wikipedia-FA: 0.675 vs. 0.730). The penalty scales with the routing failure. Proportionally, the loss is $\approx8\%$ of A3's score on Wikipedia-FA and $\approx12\%$ on QASPER --- larger where section recall is worse --- even though QASPER's absolute delta is smaller (its whole scoreboard is $\approx 3 \times$ lower).

One qualification keeps this negative honest: it is largely a no-reranker phenomenon. Under \emph{bge-reranker-v2-m3}, the QASPER gap shrinks from -0.0147 to -0.0038 (Wikipedia-FA: -0.0326 → -0.0273). The reranker re-scores a wide flat candidate list that routing would otherwise have pruned. The correct reading is therefore narrow. Route-to-section-then-rank under-recalls as the section inventory grows. It loses to flat search over the same index unless a reranker restores the candidates. This is a result about one widely-used naive architecture, not about hierarchy in general. It is consistent with the literature's split verdict \cite{laitenberger2025stronger, liu-etal-2021-dense-hierarchical, singh2025hierarchical}.

\subsection{Gold vs. induced structure is corpus-dependent}
\label{sec:corpus-struct}

On Wikipedia-FA, native heading trees beat Gemma's induced trees by +0.0129 (Holm-significant). Imperfect induction has a small, measurable cost. On QASPER the advantage vanishes (+0.0011, n.s.). This is not because the inducer copied the papers' sectioning: span identity is 75.8\%, comparable to Wikipedia's 78.9\%. Both readings are findings. For the gold-structure story: the advantage of native structure does not generalize across genres, and no one should claim it corpus-generally. For the induction story: a 26B open-weights model re-segments a quarter of all spans and loses nothing measurable against author-written structure on scientific papers. A plausible mechanism --- untested, offered as hypothesis --- is depth headroom. 61\% of QASPER papers are depth-1, so gold and induced trees have little room to differ in the information the breadcrumb carries. Wikipedia-FA's $\geq3$-deep trees give gold structure more leverage.

\subsection{Exploratory analyses}

\textbf{Breadcrumb depth dose-response.} We regress cov-nDCG@10 on injected path depth (0→3, i.e. A2 → A3-d1 → A3-d2 → A3). The strong embedder shows a small positive slope on both corpora (Wikipedia-FA +0.0032/level, bootstrap p = 0.02; QASPER +0.0027/level, p < 0.001). The weak embedder is flat on Wikipedia-FA (+0.0006/level, p = 0.65) and marginal on QASPER (+0.0012/level, p = 0.052). In every cell, most of the benefit arrives at the first path level and then plateaus (e.g., Wikipedia-FA strong: 0.403 → 0.412 → 0.413 → 0.413). We do not promote this to a finding. It sits outside the pre-registered family. The weak-embedder slopes disagree across corpora. QASPER's shallow native depth also partly collapses the depth arms by construction.

\textbf{Per-query-type cuts.} On Wikipedia-FA, the C1 gain concentrates on factoid (+0.046) and multihop (+0.031) queries. It is absent on structural queries (-0.002). C2 is largest on structural queries (+0.023). On QASPER, C1, C2, and C4 hold their signs across all three native question types (extractive/abstractive/boolean). The two corpora use incommensurable query taxonomies, so no specific type interaction can be called replicated. We report these cuts as hypothesis-generating only.

\textbf{Reranking lifts all boats.} A cross-encoder adds +0.04-0.06 cov-nDCG to essentially every condition on both corpora. It leaves the organization contrasts intact (Wikipedia-FA A3: 0.413 → 0.474; QASPER: 0.125 → 0.165). It disproportionately rescues only A4 (Section \ref{sec:naivehurts}).

\textbf{Recall, not just ranking.} The contrasts partly live in recall. On QASPER, strong-embedder recall@50 spans 0.248 (A0) to 0.302 (A3). Structure-aligned chunking retrieves more of the evidence, not merely in better order. Absolute recall ($\approx0.25–0.30$) shows this benchmark is far from saturated.

\section{Discussion}

\textbf{A mechanism ranking, for one well-specified regime.} For dense retrieval over long structured documents with strong open-weights embedders, the mechanisms rank as follows. Boundary placement comes first, with the largest and most reliable gain. Native heading-path metadata comes second: small, real, and content-driven. LLM per-chunk contextualization comes third: positive against raw windows, negative against structural cutting, and budget-expensive. Naive two-stage routing comes last: harmful without a reranker. For a practitioner, the actionable form is:

\begin{enumerate}
    \item Cut where the author cut.
    \item Prepend one level of real heading path.
    \item Spend the LLM budget you would spend on per-chunk contexts on structure induction instead. Induction needs only ~26B-class capability and matched gold structure on one of two corpora.
    \item Do not route-to-section without a reranker behind it.
\end{enumerate}

\textbf{What the controls changed.} Each methodological element did visible work. The size control is why the largest effect in the study (A0→A2) is attributable to boundary placement rather than chunk length. The placebo is why the metadata effect is attributable to content rather than token injection. The placebo also caught the null case: on QASPER, semantically null paths actively hurt relative to no prefix, an outcome no bare-baseline comparison could surface. The fixed primary cell is why the depth dose-response and query-type patterns remain labeled exploratory instead of headline claims. Cross-corpus replication demoted one of the three Wikipedia-FA "organization" pillars (C3) to a boundary condition. With one corpus we would have gotten that conclusion wrong, in a literature where single-corpus structure claims are the norm. We take this as direct evidence for the paper's methodological thesis. In a subfield with documented non-replication, mechanism claims need controls of roughly this strength before they are worth acting on.

\textbf{Reconciling the literature.} Our results locate several published disagreements. The chunking literature's instability \cite{qu-etal-2025-semantic} is consistent with our small effect sizes. Gains of +0.01–0.02 under tight controls will oscillate through zero in uncontrolled comparisons. Contextualization's mixed record \cite{merola2026reconstructing,conti2025context} matches A1's position: real against raw windows, dominated by structural cutting. The hierarchical-retrieval split \cite{liu-etal-2021-dense-hierarchical} vs. \cite{laitenberger2025stronger} resolves, in our data, into a statement about first-stage recall as a function of section inventory. That statement predicts both the successes (small inventories, strong routers) and the failures (large inventories, centroid routers).

\section{Limitations}

\textbf{Small effects.} The replicated organization effects are +0.01-0.02 cov-nDCG@10 ($dz \leq 0.11$). Applications where retrieval quality is not the bottleneck will not notice them. We claim reliability, not magnitude.

\textbf{One inducer.} Both corpora share one structure-induction model (\emph{Gemma-4-26B-A4B}). Induction quality plausibly varies with the inducing model. That would move C3 in particular, whose corpus-dependence already shows the gold/induced gap is sensitive to conditions. A second-inducer replication is the clear next experiment.

\textbf{Two corpora, one language.} Wikipedia-FA and QASPER differ usefully in genre, query provenance, and structure depth. But both are English and expository. The C3 reversal between them is itself a warning against generalization of any structure result from $\leq2$ corpora.

\textbf{Synthetic queries on one corpus.} Wikipedia-FA queries are LLM-generated, though leakage-filtered, paraphrased, and provenance-controlled. QASPER's native questions mitigate but do not eliminate the concern, because the corpora differ in more than query provenance.

\textbf{Secondary model specification.} The MixedLM omits a query random effect in a within-query paired design. We report it for pre-registration fidelity and treat the document-clustered bootstrap as primary. A qid-random-effect respecification is planned.

\textbf{One routing architecture.} C4 indicts centroid-based route-to-section-then-rank with top-5 sections — the naive but common design. It does not indict hierarchical retrieval broadly. Parent–child reranking schemes \cite{singh2025hierarchical} are untested here.

\textbf{Retrieval only.} No answer-generation stage is evaluated. Whether +0.01–0.02 cov-nDCG@10 moves downstream answer quality is out of scope.

\section{Conclusion}

We asked which of four document-structure mechanisms actually helps dense retrieval. We built the controls the answer needs to be trustworthy: size-matched baselines, a semantically null placebo, a coverage-aware primary metric, pre-registered contrasts under clustered inference, and replication across two distant corpora with different query provenances. The answer is consistent across both corpora. Organization helps, modestly. Boundary placement helps most. The content of native heading paths helps secondarily. LLM chunk contextualization underperforms structural cutting. Naive hierarchical routing hurts in proportion to its first-stage recall failure. The value of gold over induced structure, by contrast, did not survive the corpus change. The analyses we left uncontrolled --- the depth dose-response and the query-type interactions --- are precisely the ones that remain unsettled. Both facts argue the same point. In retrieval-mechanism research, the framing choice between methodology and mechanism is false. The mechanisms can only be seen through the methodology.


\section*{Ethics and Privacy Statement}

This study evaluates retrieval only. It runs no answer-generation stage and
makes no deployment claim. Both corpora are public research datasets:
Wikipedia Featured Articles and QASPER \citep{dasigi2021dataset}. Neither
contains personal data collected for this work, and the study involved no
human subjects. All LLM stages reported in this paper ran on a local
open-weights model.

The findings are small in magnitude, and we report them as such. The
practical risk we see is overreading: a +0.01 to +0.02 cov-nDCG@10 gain is
not a license to spend LLM inference on structure induction without
measuring the downstream effect on a target corpus. The negative result
carries the same caution in reverse. It indicts one common routing design,
not hierarchical retrieval as a family.

\bibliographystyle{acl_natbib}
\bibliography{base}

\appendix
\clearpage

This appendix documents the experimental instrument in full. Section A names the eleven conditions and the grid that holds them. Sections B and C define the primary metric and its relatives. Section D states the statistical protocol as the code runs it. Sections E to G build both corpora and state the ground-truth checks. Section H covers structure induction. Section I covers the non-lexical query sets. Section J covers the reproducibility envelope. When this appendix reports an implementation detail, it names the module and the behavior. The released code is the final authority.

\section{Conditions, naming conventions, and the experimental grid}

\subsection{The two condition families}

Every condition is one point in a design that varies exactly three things. First, where the chunk boundaries fall. Second, what prefix precedes the chunk content at embedding time. Third, whether retrieval is flat or two-stage. Chunks are contiguous, non-overlapping spans that tile each document exactly. One run holds all other factors fixed: corpus, embedder, and retrieval budget.

Condition names encode two axes. The letter encodes the source of structure:
\begin{itemize}
    \item \textbf{\emph{A} means induced structure.} The local Gemma-4-26B-A4B model produces the section tree after the pipeline strips the native headings (See appendix \ref{apx:induction}).
    \item \textbf{\emph{G} means gold structure.} The section tree is the document's own author-written heading tree, offset-verified (See appendix \ref{apx:offset}).
\end{itemize}
The number encodes the treatment level. The numbering stays parallel between the two families. A G-condition is therefore the exact counterpart of the A-condition with the same number. The two differ only in gold structure versus induced structure:

\begin{table*}[t]
\centering
\small
\caption{Condition levels.}
\label{tab:levels}
\begin{tabular}{@{}clcc@{}}
\toprule
\textbf{Level} & \textbf{Meaning} & \textbf{Induced} & \textbf{Gold} \\
\midrule
0 & fixed-window chunks (size-matched), no prefix & A0 & --- \\
1 & fixed-window chunks + LLM situating context ($\leq$80 tok) & A1 & --- \\
2 & structural chunks, no prefix & A2 & G2 \\
3 & structural chunks + heading-path prefix (depth 3) & A3 & G3 \\
4 & structural chunks + heading-path prefix + two-stage routing & A4 & G4 \\
\bottomrule
\end{tabular}
\end{table*}

There is no G0 or G1, by design. Levels 0 and 1 use fixed windows and therefore have no structure. "Gold versus induced structure" is undefined for them. The G family exists only where a structural tree is in use (levels 2 to 4).

\subsection{The placebo and the depth ladder}

Two conditions extend the induced family for control purposes:
\begin{itemize}
    \item \textbf{P (placebo)} matches A3 (induced structure, depth-3 heading-path prefix, flat). It differs in one way: each chunk's prefix is a real heading path from a different document. The prefix matches the real ones in format, depth, and length distribution. It carries no information about the chunk. P isolates the token-injection confound (main text \ref{sec:controls}; construction in \ref{sec:placebo} here).
    \item \textbf{A3-d1 / A3-d2 / A3} form the breadcrumb depth ladder. The \verb|-dN| suffix means "heading path cut to depth N." \textbf{The bare name A3 is the full-depth (depth-3) rung.} The "3" in "A3" is the treatment-level index. It coincides with the depth, which is why the name is not \verb|A3-d3|. Read as a dose-response series, the ladder is \verb|A2| (depth 0, no prefix), then \verb|A3-d1|, then \verb|A3-d2|, then \verb|A3| (depth 3). The prefix renders as \verb|[Title › Section › … ]|\discretionary{}{}{}\verb| \n\n content|. The separator is \verb|›|.
\end{itemize}

\subsection{The eleven conditions}

\begin{table*}[t]
\centering
\small
\begin{tabular}{@{}lllll@{}}
\toprule
\textbf{Condition} & \textbf{Boundaries} & \textbf{Prefix} & \textbf{Retrieval} & \textbf{Structure} \\
\midrule
A0 & fixed windows (size-matched) & none & flat & --- \\
A1 & fixed windows (size-matched) & LLM situating context ($\leq$80 tok) & flat & --- \\
A2 & induced structure & none & flat & induced \\
A3-d1 & induced structure & heading path, depth 1 & flat & induced \\
A3-d2 & induced structure & heading path, depth 2 & flat & induced \\
A3 & induced structure & heading path, depth 3 & flat & induced \\
P & induced structure & \textit{shuffled} heading path, depth 3 & flat & induced \\
A4 & induced structure & heading path, depth 3 & two-stage & induced \\
G2 & gold structure & none & flat & gold \\
G3 & gold structure & heading path, depth 3 & flat & gold \\
G4 & gold structure & heading path, depth 3 & two-stage & gold \\
\bottomrule
\end{tabular}
\caption{Experimental conditions.}
\label{tab:conditions-full}
\end{table*}

\subsection{Contrasts, and the cell that tests them}

Each pre-registered contrast is a single subtraction that isolates one mechanism:

\begin{table*}[t]
\centering
\small
\begin{tabular}{@{}llp{8cm}@{}}
\toprule
\textbf{Contrast} & \textbf{Subtraction} & \textbf{Question} \\
\midrule
C1 & A3 $-$ A1 & Do structure-aligned boundaries plus native metadata beat the strongest generative treatment on fixed windows? \\
C2 & A3 $-$ P & Does the \textit{content} of the heading path matter, or is the gain token injection? \\
C3 & G3 $-$ A3 & How much is lost when an LLM induces the structure instead of the author? \\
C4 & A4 $-$ A3 & Does two-stage hierarchical retrieval beat flat search over the identical index? \\
\bottomrule
\end{tabular}
\caption{Pre-registered contrasts.}
\label{tab:contrast-questions}
\end{table*}

A \textbf{cell} is one fixed combination of the non-condition factors: one metric, one embedder, one reranker setting, one corpus. The full experiment is a grid over those factors. The conditions above re-run in each cell. One pre-registered primary cell gives the four contrasts their confirmatory test, which forecloses metric-shopping: \emph{cov-nDCG@10 × strong embedder (bge-large) × no reranker}. Every other cell is exploratory, and main-text \ref{sec:results} labels it as such. This covers other metrics, the weak embedder, and the reranked variants. The grid axes are these. Metric is one of cov-nDCG@10, recall@50, MRR, binary-nDCG@10, or token-budget coverage at 512/1024/2048. Embedder is the weak bge-small-en-v1.5 or the strong bge-large-en-v1.5. Reranker is none or bge-reranker-v2-m3 over the top 50. Corpus is Wikipedia-FA or QASPER.

\section{Coverage-nDCG@10: formal definition}

\subsection{Motivation}

The evidence for a single query often spreads across several disjoint character spans. On QASPER, 43.9\% of questions cite multiple evidence paragraphs (mean 2.10 spans/question). The primary metric therefore needs two properties. Binary nDCG provides neither:
\begin{enumerate}
    \item \textbf{Novelty (coverage) awareness.} A chunk that repeats evidence already retrieved higher in the ranking should earn little. Credit accrues for new gold coverage only. Without this property, a condition that cuts text into many small chunks banks the same span again and again.
    \item \textbf{Per-condition ideal normalization.} A condition is scored against the best ranking that its own chunk set allows. A condition is therefore not penalized for segmenting the document differently from another condition.
\end{enumerate}
Coverage-nDCG applies the novelty principle of $\alpha$-nDCG \cite{clarke2008novelty} one level below sub-topic diversity. It applies it at the level of gold-evidence characters within a single query.

\subsection{Definition}

Let a query belong to document $d$ with gold-evidence spans $G = \{g_1, \dots, g_S\}$. Each $g_j = [\text{start}_j, \text{end}_j)$ is a half-open character interval on $d$'s flat text, and $|g_j|$ is its length. A retrieval run returns a ranked list of chunks $c_1, c_2, \dots$. A chunk from any document other than $d$ contributes zero, because offsets are per-document coordinate spaces.

For each gold span $g_j$, keep the set of its characters already covered by chunks ranked above the current one. For the chunk $c_i$ at rank $i$, the \textbf{marginal gain} is the mean over gold spans of the fraction of newly covered characters:

$$
\text{gain}(c_i) \;=\; \frac{1}{S} \sum_{j=1}^{S} \frac{\bigl|\, (c_i \cap g_j) \setminus \text{covered}_{<i}(g_j) \,\bigr|}{|g_j|}
$$

Here $\text{covered}_{<i}(g_j)$ is the union of the portions of $g_j$ covered by $c_1, \dots, c_{i-1}$. After the metric scores $c_i$, it adds the overlap of $c_i$ with each $g_j$ to $\text{covered}(g_j)$. 

The discounted cumulative gain at cutoff $k$ uses the standard log discount:

$$
\text{DCG@}k \;=\; \sum_{i=1}^{k} \frac{\text{gain}(c_i)}{\log_2(i+1)}
$$

The ideal DCG runs over the same condition's own chunk set by a greedy maximum-marginal-coverage ordering. It repeatedly appends the not-yet-selected chunk with the largest current marginal gain, up to $k$ chunks. This greedy order is the standard tractable stand-in for the ideal ranking under a coverage objective. Finally,

$$
\text{cov-nDCG@}k \;=\; \frac{\text{DCG@}k}{\text{IDCG@}k}, \qquad k = 10.
$$

The value is 0 when $\text{IDCG@}k = 0$. That case means no chunk of the condition covers any gold character.

\subsection{Worked micro-example}

Suppose one query has a single gold span $g_1$ of length 100 characters. Condition X returns, in rank order, chunks that cover characters $[0,60)$, then $[40,100)$, then an irrelevant chunk:

\begin{itemize}
    \item Rank 1 covers 60 new chars. The gain is $60/100 = 0.60$. The discount is $\log_2 2 = 1$.
    \item Rank 2 covers $[40,100)$, but $[40,60)$ is already covered, so 40 new chars. The gain is $0.40$. The discount is $\log_2 3 \approx 1.585$.
    \item Rank 3 covers 0 new chars. The gain is 0.
\end{itemize}

The result is $\text{DCG} = 0.60/1 + 0.40/1.585 = 0.852$. The greedy ideal from this chunk set places the same two chunks first, because their gains are identical. So $\text{IDCG} = 0.852$ and $\text{cov-nDCG} = 1.0$. Suppose instead a third chunk covered $[0,100)$ in one shot. The ideal would rank it first (gain 1.0 at rank 1), and the observed ranking would score below 1.0. The example shows the two active ingredients. The rank-2 chunk is discounted for redundancy (40 new chars, not 60). The score is relative to the best that the condition's own chunks allow.

\section{Relevance rule and secondary metrics}
\label{sec:relevance}

Coverage-nDCG uses graded character coverage and never thresholds. The \textbf{binary} relevance rule below serves only the secondary metrics that need a yes/no judgment.

\textbf{Binary relevance}. A chunk $c$ is relevant to a gold span $g$ if

$$
\frac{|c \cap g|}{|g|} \ge 0.5 \quad\text{(the chunk substantially contains the evidence)} \qquad\text{or} 
$$
$$
\frac{|c \cap g|}{|c|} \ge 0.8 \quad\text{(the chunk lies substantially inside the evidence)}.
$$

All overlap runs on \textbf{content-only} character offsets. A prefix therefore can never create spurious relevance.

\textbf{Secondary metrics} form a ladder from rank-insensitive to application-realistic:

\begin{itemize}
    \item \textbf{recall@50} is the fraction of gold spans that the union of the top-50 chunks covers by $\ge50\%$.
    \item \textbf{MRR} is the reciprocal rank of the first binary-relevant chunk.
    \item \textbf{binary-nDCG@10} is nDCG with 0/1 gains under the binary rule. It normalizes by the ideal binary ranking of the condition's own chunk set.
    \item \textbf{token-budget coverage} at 512/1024/2048 tokens is the gold coverage that top-ranked chunks reach until the budget runs out. \textbf{Prefix tokens count against the budget.} Each treatment therefore pays its real context-window cost. This is where contextualization's cost becomes explicit. Main-text \ref{sec:wherecut} shows that at 512 tokens A1 falls below plain A0 on both corpora.
\end{itemize}

\section{Statistical protocol}

\subsection{Paired per-query deltas}

Each contrast runs on \textbf{paired per-query deltas}. For every query answered under both conditions, $\delta_q = \text{score}_A(q) - \text{score}_B(q)$. The reported effect is the unweighted mean $\bar\delta$ over queries. Pairing is the correct design, because every query is scored under both arms. Pairing removes the large query-to-query difficulty variance from the comparison.

\subsection{Cluster bootstrap (primary)}

Queries are not independent. Several queries share a document and move together. The protocol therefore estimates uncertainty with a document-clustered bootstrap, not an ordinary per-query bootstrap. The difference is the resampling unit:

\begin{itemize}
    \item Group the per-query deltas by their document.
    \item For each of $B=10,000$ iterations, resample documents with replacement. Every query of a drawn document enters the replicate as a bundle.
    \item Recompute the mean delta on each replicate: $\bar\delta^{(b)} = \big(\sum_{d \in \text{draw}} \text{sum}_d\big) / \big(\sum_{d \in \text{draw}} \text{count}_d\big)$.
\end{itemize}

The bootstrap distribution gives three quantities:

\begin{itemize}
    \item The \textbf{95\% CI} is the 2.5th and 97.5th percentiles of $\{\bar\delta^{(b)}\}$.
    \item The \textbf{two-sided p-value} is $2 \cdot \min\big(\Pr[\bar\delta^{(b)} \le 0],\ \Pr[\bar\delta^{(b)} \ge 0]\big)$, clipped to 1.
    \item The **effect size** is Cohen's $dz = \bar\delta / \operatorname{sd}(\delta_q)$. The standard deviation runs over the per-query deltas (ddof = 1).
\end{itemize}

Resampling documents rather than queries widens the intervals. The wider intervals reflect the true number of independent units, which is documents, not queries. Take QASPER's design: 4,303 queries over 1,508 documents. An ordinary bootstrap would treat the effective sample size as about $3\times$ too large and understate uncertainty. The cluster bootstrap is the correctly specified test, and it is primary throughout.

\subsection{Holm correction}

The protocol tests four contrasts as a family. It therefore adjusts the per-test p-values by the \textbf{Holm step-down} procedure. Sort the four p-values in ascending order. Inflate the $i$-th smallest (0-indexed) by the factor 
$(m-i)$ with $m=4$, under a running maximum so the adjusted values stay monotone. Clip to 1. Holm controls the family-wise error rate at 0.05. It rejects more true effects than the blunt Bonferroni factor of 4.

\subsection{MixedLM (secondary) and its known misspecification}

The protocol reports a mixed-effects regression as a \textbf{pre-registered secondary check}. This model is misspecified for the design, by design, and the paper discloses this. The study is paired \textit{within} query, because the same query appears under both conditions. But the model has \textbf{no per-query random effect}. The large query-to-query baseline variance therefore lands in the residual and inflates the standard error of the treatment coefficient. The model is therefore \textbf{conservative for small paired effects}. It resolves only the largest effect. This matches the observed pattern. On Wikipedia-FA ($n \approx 951$ per condition) it flags only C4 ($|dz| = 0.17$, roughly twice the others) and misses C1/C2/C3 ($dz \approx 0.08–0.10$). At QASPER's n = 4,303 the larger sample overwhelms the inflated errors, and it confirms all three non-null contrasts.

The disagreements between the bootstrap and the MixedLM on the small Wikipedia-FA effects are therefore a specification artifact of the secondary model, not evidence that the effects are fragile. The bootstrap is correctly specified and primary. The paper reports the MixedLM for pre-registration fidelity only.

\section{Wikipedia Featured Articles: corpus construction}

The pipeline samples documents from Wikipedia's Featured Articles under hard quality filters, with a fixed RNG seed (\verb|seed = 20260710|):

\begin{table*}[t]
\centering
\small
\begin{tabular}{@{}llp{6cm}@{}}
\toprule
\textbf{Filter} & \textbf{Threshold} & \textbf{Rationale} \\
\midrule
\texttt{min\_words} & $\geq$ 3{,}000 & long documents where structure can matter \\
\texttt{min\_depth} & $\geq$ 3 & heading depth $\geq$ 3 , so a non-trivial tree exists \\
\texttt{min\_sections} & $\geq$ 12 & enough sections to route among / cut at \\
\texttt{n\_docs} & 200 & scale-up target \\
\bottomrule
\end{tabular}
\caption{Document filtering criteria.}
\label{tab:filters}
\end{table*}

The pipeline fetches the FA title list, shuffles it under the seed, and walks it. It fetches each candidate, flattens it with the single canonical normalizer, and accepts it if it passes the filters and the per-topic cap (\ref{sec:topicdistrib}). The loop stops at 200 accepted documents.

\subsection{Topic distribution: a per-topic cap}
\label{sec:topicdistrib}

The classifier assigns each accepted document to one of six keyword-defined topic buckets. It matches the document's non-hidden Wikipedia categories against a fixed keyword table. The first matching bucket in table order wins. An other bucket is the fallback:

\begin{table*}[t]
\centering
\small
\begin{tabular}{@{}lp{9cm}@{}}
\toprule
\textbf{Bucket} & \textbf{Example keywords} \\
\midrule
\texttt{history\_war} & battle, war, military, campaign, siege, history, empire, dynasty \\
\texttt{biography} & births, deaths, people, biograph, politicians, writers, actors, monarchs \\
\texttt{science\_nature} & species, birds, mammals, physics, chemistry, biology, astronom, geolog \\
\texttt{arts\_media} & albums, songs, films, novels, paintings, opera, television, video games \\
\texttt{geo\_places} & geography, cities, rivers, mountains, roads, buildings, hurricanes, storms \\
\texttt{other} & (fallback when nothing matches) \\
\bottomrule
\end{tabular}
\caption{Topic buckets and example keywords.}
\label{tab:buckets}
\end{table*}

The "balancing" is a single \textbf{hard ceiling}: \texttt{max\_per\_bucket = 60}. The seeded walk enforces it. It rejects a candidate whose bucket already holds 60 accepted documents. There is no minimum-per-bucket guarantee and no round-robin. The cap only limits over-represented topics. It does not equalize them.

\begin{table}[tbp]
\centering
\small
\caption{Realized distribution, verified 2026-08-04.}
\label{tab:realized-distribution}
\begin{tabular}{@{}lrr@{}}
\toprule
\textbf{Bucket} & \textbf{Documents} & \textbf{Share} \\
\midrule
history\_war    & 60 \textit{(at cap)} & 30.0\% \\
other           & 60 \textit{(at cap)} & 30.0\% \\
biography       & 30 & 15.0\% \\
arts\_media     & 22 & 11.0\% \\
science\_nature & 15 & 7.5\% \\
geo\_places     & 13 & 6.5\% \\
\midrule
\textbf{Total} & \textbf{200} & 100\% \\
\bottomrule
\end{tabular}
\end{table}

The cap binds on the two largest buckets. Both land exactly at 60. Without it, \texttt{history\_war} and \texttt{other} would each be larger. 60\% of the corpus still sits in two buckets. This reflects the genuine topic skew of Featured Articles. The paper discloses this skew rather than hides it, and the skew strengthens the generalization claim. The replicated organization effects hold over a topic-imbalanced corpus, not a curated one.

\subsection{Offset-verified gold structure}
\label{apx:offset}

Gold structure is the article's own heading tree. Two layers make its character offsets trustworthy.

\textbf{Exact by construction.} The pipeline builds the flat document string paragraph by paragraph. It records each paragraph's as it emits the paragraph. It reads gold section offsets directly from this paragraph table. It never recovers them by searching for a heading's text. A text search is the step that would introduce off-by-N drift.

\textbf{Mechanically verified.} Segment verification returns a list of violations. An empty list means verified. It enforces that the sections:

\begin{itemize}
    \item \textbf{tile the document with no gaps or overlaps.} Each segment starts exactly where the previous one ended. The final segment ends at the last paragraph.
    \item have \textbf{offsets that agree with the paragraph table} exactly. Otherwise it reports "char offsets disagree with paragraph table".
    \item have in-bounds paragraph ranges and legal nesting levels ($1 \leq level \leq MAX\_LEVEL=4$).
\end{itemize}

Relevance runs purely as character-span overlap (\ref{sec:relevance}). These offsets are therefore load-bearing. An off-by-N error would silently corrupt every relevance judgment. The same offset discipline applies to induced structure (\ref{sec:structure}). Induced structure emits paragraph index ranges rather than raw offsets. This makes "structure without content modification" a verified property and holds the gold and induced arms to one integrity standard.

\subsection{Placebo construction}
\label{sec:placebo}

Placebo prefixes are real heading paths from other documents. The pipeline buckets unique \verb|(document, path)| pairs by depth. It permutes them within depth-bucket toward a cross-document derangement under a fixed seed. It counts and reports residual same-document assignments. The result matches real breadcrumbs in format, depth, and length distribution. It carries no information about the chunk it attaches to. This is a semantically null treatment in the strict sense.

\section{Wikipedia-FA query generation and leakage control}
\label{sec:wikipedia}

Queries are LLM-generated but provenance- and leakage-controlled. The pipeline itself samples the gold spans. The query-writing model never self-reports them.

\textbf{Four query types} per document, over sampled gold spans:

\begin{itemize}
    \item \textbf{factoid} targets one span.
    \item \textbf{underspecified} targets one span but names the referent only vaguely ("the commander", never the proper name).
    \item \textbf{multihop} targets two spans in different top-level sections.
    \item \textbf{structural} targets a whole top-level section.
\end{itemize}

\textbf{Two-stage leakage filter.} Each generated query must pass a lexical-leakage filter. The filter rejects it if ROUGE-L F1 against its gold span exceeds 0.35, or if it shares any verbatim 4-gram with the span. A second LLM call then paraphrases the query. The paraphrase enters the benchmark, and it must pass the same filter again.

\textbf{Funnel:} 2,306 generated → 1,945 survivors → 951 balanced finals (250 factoid, 250 underspecified, 250 structural, 201 multihop) over 184 documents. Shipped queries stay lexically far from their evidence (typical ROUGE-L 0.02–0.13).

\textbf{Non-lexicality check.} BM25 places a gold chunk first for only 36.3\% of the shipped queries. This is direct evidence that the leakage controls worked and that keyword overlap does not solve the benchmark (see \ref{sec:nonlex}).

\section{QASPER: corpus construction and evidence mapping}

QASPER \cite{dasigi2021dataset} supplies NLP papers with questions from real readers and gold evidence \textit{paragraphs}. This is human-authored relevance on a genre far from encyclopedic prose. The study is an IR ablation and not supervised learning, so the pipeline merges all splits. Gold structure is each paper's own section tree.

\textbf{Evidence mapping.} A cascade matches each question's gold evidence paragraphs into the flattened document: exact match, then fuzzy match, then section fallback. 95.7\% of evidence paragraphs match exactly.

\textbf{Filtering funnel:} 5,049 → 4,303 usable questions over 1,508 papers, after the pipeline drops unanswerable, no-evidence, and numeric-only items.

\textbf{Multi-span profile.} 43.9\% of questions cite multiple evidence paragraphs (mean 2.10 spans/question). This is the regime that coverage-nDCG exists for.

\textbf{Difficulty and non-lexicality.} BM25 gold-top-1 is 22.7\%. This set is harder and more non-lexical than the Wikipedia-FA set (see \ref{sec:nonlex}).

\textbf{Shallow native depth.} 61\% of QASPER papers have depth-1 section lists. Every Wikipedia-FA document has depth $\geq 3$ by construction. Gold and induced trees therefore have less room to diverge in the information a breadcrumb can carry. This is the offered explanation, untested, for the null C3 on QASPER against the positive C3 on Wikipedia-FA (main-text \ref{sec:corpus-struct}).

\section{Structure induction protocol}
\label{sec:structure}
\label{apx:induction}

The induced-structure conditions (A2–A4, P) get their hierarchy from the local \textbf{Gemma-4-26B-A4B} model.

\textbf{Input and output.} The model sees the document's paragraphs with all native headings stripped. It returns sections as \verb|(title, level, paragraph-range)| tuples under hard constraints. The sections must \textbf{tile the paragraphs exactly}. The levels may increase by at most one between consecutive sections. The model emits paragraph index ranges rather than character offsets. This makes "structure induction without content modification" a mechanically verifiable property.

\textbf{Validation and repair.} The pipeline checks the constraints mechanically. A failing output gets \textbf{one self-correction round} that quotes the specific violations back. Then a \textbf{deterministic repair} snaps a near-valid tiling to a legal one. The pipeline clamps nesting deeper than \verb|MAX_LEVEL = 4|, so the gold and induced arms share one level budget. The clamp bounds the stored level integer only. The true depth still drives the breadcrumb path.

\begin{table*}[t]
\centering
\small
\begin{tabular}{@{}lccc@{}}
\toprule
\textbf{Corpus} & \textbf{Usable structure} & \textbf{Required repair} & \textbf{Mean sections/doc} \\
\midrule
Wikipedia-FA & 198 / 200 & 116 & 21.9 \\
QASPER & 1{,}579 / 1{,}585 & 590 & 14.5 (gold 14.8) \\
\bottomrule
\end{tabular}
\caption{Corpus structure statistics.}
\label{tab:corpus-structure}
\end{table*}

\textbf{The inducer does not merely recover native sectioning.} Only 78.9\% (Wikipedia-FA) and 75.8\% (QASPER) of induced chunk spans are byte-identical to a gold span. Gold and induced conditions therefore measure genuinely different segmentations. This is what makes C3 (G3 − A3) a substantive comparison rather than a tautology.

\section{Non-lexicality and the BM25 baseline}
\label{sec:nonlex}

\textbf{What it measures.} BM25 is the classic lexical (keyword-overlap) retriever. "BM25 gold-top-1" is the fraction of queries for which BM25 ranks a gold-relevant chunk in position 1. This is a direct probe of how solvable the benchmark is by word matching alone.

\begin{table}[tbp]
\centering
\small
\begin{tabular}{@{}lc@{}}
\toprule
\textbf{Corpus} & \textbf{BM25 gold-top-1} \\
\midrule
Wikipedia-FA & 36.3\% \\
QASPER & 22.7\% \\
\bottomrule
\end{tabular}
\caption{BM25 gold-top-1 retrieval rate.}
\label{tab:bm25-gold-top1}
\end{table}

Both are low, and QASPER is the harder of the two. A keyword search puts a correct chunk first for under a quarter of its questions. Three consequences matter for this study:

\begin{itemize}
    \item \textbf{Headroom.} The benchmarks are far from lexically saturated. A meaning-based (dense) retriever therefore has room to matter, and structure has room to help it.
    \item \textbf{Regime.} The paper's claims are about dense retrieval and how document structure helps it. A non-lexical benchmark forces the retriever into exactly that regime. The gains therefore cannot be dismissed as trivial keyword effects.
    \item \textbf{Leakage defense} (Wikipedia-FA especially). For LLM-generated queries, the sharpest objection is that the generator copied words from the evidence. The low BM25 hit rate is the empirical proof that the ROUGE-L/4-gram filters and the paraphrase step (\ref{sec:wikipedia}) worked. The shipped queries really are lexically distant from their evidence. QASPER's questions are human-written, so its 22.7\% is naturally occurring hardness. Two genres, two query provenances, both genuinely non-lexical.
\end{itemize}

\section{Reproducibility and configuration}

\textbf{Determinism.} The measured half of the pipeline is deterministic: embedding, retrieval, scoring, statistics. All LLM generation stays in one-time corpus and query construction under fixed seeds with on-disk caching. This covers structure induction, A1 contexts, query generation, and paraphrase. Generation therefore does not enter the measured loop. All generative stages ran on a local \emph{Gemma-4-26B-A4B} model via \emph{vLLM} on a single H200 GPU. Table \ref{tab:config} lists reproducibility parameters used in the study.

\begin{table*}[t]
\centering
\small
\caption{Key configuration (\texttt{configs/scale\_crc.yaml}).}
\label{tab:config}
\resizebox{\textwidth}{!}{%
\begin{tabular}{@{}llp{6cm}@{}}
\toprule
\textbf{Group} & \textbf{Parameter} & \textbf{Value} \\
\midrule
global & \texttt{seed} & 20260710 \\
selection & \texttt{n\_docs} / \texttt{min\_words} / \texttt{min\_depth} / \texttt{min\_sections} / \texttt{max\_per\_bucket} & 200 / 3000 / 3 / 12 / 60 \\
chunking & \texttt{t\_max} / \texttt{t\_min} (content tokens) / \texttt{fixed\_snap\_tokens} & 350 / 60 / 15 \\
breadcrumb & \texttt{sep} / \texttt{depths} & \texttt{"\ >\ "} / [1, 2, 3] \\
querygen & \texttt{per\_type} / \texttt{rouge\_max} / \texttt{ngram\_block} / \texttt{context\_max\_tokens} & 250 / 0.35 / 4 / 80 \\
embedders & weak / strong & bge-small-en-v1.5 / bge-large-en-v1.5 \\
reranker & model / top / arm & bge-reranker-v2-m3 / 50 / strong \\
retrieval & \texttt{top\_k} / \texttt{hier\_top\_sections} & 100 / 5 \\
metrics & \texttt{ndcg\_k} / \texttt{recall\_k} / \texttt{budgets} & 10 / 50 / [512, 1024, 2048] \\
stats & \texttt{bootstrap\_iters} & 10{,}000 \\
\bottomrule
\end{tabular}%
}
\end{table*}

\textbf{Size matching (guards C1).} The fixed-window baseline's per-document target window equals the mean token length of that document's induced chunks. Realized means differ by $\leq4$ content tokens per chunkset (Wikipedia-FA: 255.4 fixed vs. 259.6 induced; QASPER: 244.9 vs. 249.8). The A0/A1-vs-A2/A3 comparison therefore isolates boundary placement, not chunk size. Window edges snap to sentence boundaries. Structural chunks split at sentence boundaries or merge when they exceed 350 or fall below 60 content tokens.

\textbf{Representation constancy (guards C4).} Two-stage conditions build section centroids as the renormalized mean of the same chunk vectors that flat retrieval searches. C4 therefore compares retrieval algorithms, never representations. Stage 1 ranks section centroids and keeps the top 5 (a section is a document plus a top-level heading). Stage 2 ranks chunks within the routed sections. If fewer than $k$ candidates remain, the system backfills from the global ranking. A per-query section-recall diagnostic records whether stage 1 retained any gold-overlapping section. This separates routing failures from ranking failures.

\textbf{Release.} The authors will release code, configurations, and per-query results.

\end{document}